# Coherent vibrational wave packet motion in *Er*Cry4a proteins monitors the redox state of the flavin chromophore

*Krishan Kumar,[a] Daniel Timmer,[a,†] Maryam Khazani,[b] Glen Dautaj,[b] Sophie Brinkmann,[b] Jan P. Götze,[c] Peter Saalfrank,[d] Boris F. Gribakin,[a] Antonietta De Sio,[a,e] Henrik Mouritsen,[b,f*] and Christoph Lienau[a,e,f*]*

[a] Institut für Physik, Carl von Ossietzky Universität Oldenburg, 26129 Oldenburg, Germany

[b] AG Neurosensory Sciences/Animal Navigation, Institut für Biologie und Umweltwissenschaften, Carl von Ossietzky Universität Oldenburg, 26129 Oldenburg, Germany

[c] Institut für Chemie und Biochemie, Freie Universität Berlin, 14195 Berlin, Germany

[d] Institute of Chemistry, University of Potsdam, 14476 Potsdam, Germany

[e] Center for Nanoscale Dynamics (CENAD), Carl von Ossietzky Universität Oldenburg, 26129 Oldenburg, Germany

[f] Research Center for Neurosensory Sciences, Carl von Ossietzky Universität Oldenburg, 26111 Oldenburg, Germany

†present address: Fachbereich Physik, Philipps-Universität Marburg, 35032 Marburg, Germany

*Correspondence to: christoph.lienau@uni-oldenburg.de, henrik.mouritsen@uni-oldenburg.de

ORCIDs

KK: 0000-0001-5466-0299;

DT: 0000-0001-7541-8047;

MK: 0009-0004-8879-549X;

GD: 0009-0001-6631-8682;

SB: 0009-0005-7243-4099;

JPG: 0000-0003-2211-2057;

PS: 0000-0002-5988-5945;

BFG: 0000-0002-5339-1813;

ADS: 0000-0003-2363-5634;

HM: 0000-0001-7082-4839;

CL: 0000-0003-3854-5025

## ABSTRACT

Cryptochromes are blue-light-sensitive flavoproteins that play central roles in biological function. In European robin (*Erithacus rubecula*) *Er*Cry4a proteins, optical excitation of their flavin chromophore forms a long-lived radical pair through a sequence of electron transfer steps across a tetradic chain of tryptophan residues, making them primary candidates for magnetoreception in night migratory songbirds. Recent quantum chemical calculations indicate that nonadiabatic couplings play a central role in the energy and charge transfer processes initiated by optical excitation. Here, we study these dynamics in *Er*Cry4a using ultrafast transient absorption spectroscopy with 10-fs time resolution in the 450-nm spectral range. We uncover a rapid, sub-50 fs red shift in stimulated emission, quenched within 360 fs by electron transfer from a nearby tryptophan moiety. While high-frequency excited state vibrations are rapidly damped, coherent motion involving several low-frequency vibrations persists during both the initial energy relaxation and the subsequent electron transfer. This is evidenced by probing the coherent vibrational motion of the formed $FAD^{\bullet-}$ radical anion and is independently validated by blocking the electron transfer through site-selective tryptophan mutation. Our results only provide insight into the role of nonadiabatic couplings for the initial steps of cryptochrome photoactivation and suggest a general strategy for redox-state-specific monitoring of charge transfer dynamics by probing coherent vibrational motion.

## 1. Introduction

Cryptochromes (Cry) are a class of flavoproteins that belong to the photolyase/cryptochrome family[1, 2] and contain a flavin adenine dinucleotide (FAD) cofactor which acts as a blue light photoreceptor.[3, 4] These proteins are known for their role in maintaining circadian rhythm across humans,[5] plants[6] and animals,[6-8] and in plant growth regulation.[9, 10] In addition, cryptochromes have been suggested to function as light-dependent, radical-pair based magnetoreceptors.[11, 12]

In most birds, three cryptochrome genes exist, *Cry1*, *Cry2*, and *Cry4,*[13] for which six isoforms have been reported so far: Cry1a, 1b, 2a, 2b, 4a, and 4b.[14-24] A cryptochrome must bind to FAD cofactor to be a candidate magnetoreceptor, because FAD absorbs the light needed for the for-mation of the magnetically sensitive radical pairs.[12, 25, 26] Cry4a is currently the only bird cryptochrome known to bind FAD[27-30] and its mRNA levels are upregulated during migratory seasons.[20] The splice variant Cry4b, on the other hand, does not seem to be relevant for magnetoreception.[31] Recombinantly expressed and purified European robin (*Erithacus rubecula*) and chicken (*Gallus gallus*) Cry4a (*Er*Cry4a and *GgCry*4a) are sensitive to external magnetic fields *in vitro.*[25, 32] At present, Cry4a is therefore the most promising primary magnetoreceptor candidate in birds and the only avian cryptochrome in which we can currently study the radical-pair mechanism in detail.

The cryptochrome 4a protein, *Er*Cry4a, from night-migratory European robins is both light-sensitive and expressed in the retina,[14, 20, 25] leading to the proposal that these proteins act as light-dependent sensors of the Earth's magnetic field in animal navigation.[11, 12, 33] Behavioral experiments support this proposal[33-35] and indicate an inclination compass which can be disrupted by red light[36], or by electromagnetic noise[37-42].

In *Er*Cry4a, blue light is absorbed by the isoalloxazine ring of its FAD cofactor. This optical excitation triggers a sequence of consecutive electron transfer steps across a tetradic chain of tryptophans (Trp) which connects FAD to the outside of the protein.[25] This electron transfer cascade leads to the formation of a long-lived spin-correlated radical pair[12, 25, 32, 43, 44] which is thought to serve as a molecular compass for the weak Earth's magnetic field (~50 μT).[12] Indeed, *in vitro* spectroscopies have monitored effects of few-millitesla magnetic fields on the radical pair concentration upon blue-light excitation of *Er*Cry4a.[25, 32]

The optical properties of *Er*Cry4a are strongly connected to their Flavin chromophore, whose photophysical properties have been thoroughly investigated using transient absorption (TA)[45-57] and time-resolved fluorescence studies.[3, 49, 58-61] TA studies gave insight into the dynamics of the dif-ferent redox states of FAD,[25, 51] existing in planar or bent isoalloxazine ring structure.[49, 55] Stimulated Raman spectroscopy has been used to characterize the vibrational modes of FAD in the excited state.[56]

Electronic structure calculations of flavin molecules show two optically bright ($\pi\pi^*$) and two dark ($n\pi^*$) states in the visible energy region.[62-67] This absorption spectrum is dominated by a 2.78 eV band (Fig. 1a), assigned to the first bright transition while the second bright transition is at 3.35 eV. Two additional optically dark states in this energy range have solvent-dependent energies.[68] Mixed quantum classical calculations on riboflavin predict that nonadiabatic couplings between these bright and dark states[66] should profoundly affect the ultrafast dynamics. A strong red shift of the stimulated emission (SE) spectra and rapid intramolecular vibrational energy redistribution (IVR) processes have been predicted, supporting conclusions from earlier TA studies.[56, 68] In recent TA experiments on FAD in water, we have resolved this transient SE red shift, occurring within only 20 fs, and the ensuing IVR dynamics.[57]

This raises the question of how the protein environment affects these nonadiabatic couplings and -more generally - which role vibronic couplings play for the electron transfer in the protein environment. Very recently, some of these questions have been raised in theoretical work, predicting nonadiabatic activation pathways in cryptochromes[69, 70] and low-lying conical intersections in lumiflavins.[71] So far, time-resolved studies of nonadiabatic couplings in cryptochromes are lacking. Previous experimental studies[43, 55, 72-74] have been performed with a time resolution of more than one vibrational cycle, making them insensitive to coherent vibrational motion in the protein environment.

Here, we investigate the transient absorption of *Er*Cry4a proteins at 4 °C with 12-fs time resolution for blue light excitation. We uncover distinct signatures of nonadiabatic couplings, specifically a pronounced red shift of stimulated emission within ~20 fs and a suppression of high-frequency excited state vibrational motion. In contrast, low-frequency (< 600 cm$^{-1}$) coherent wavepacket motion persists beyond the initial electron transfer step. We find mode-selective, spectrally surprisingly narrow excited state absorption spectra of the semiquinone flavin (FAD$^{\bullet-}$) radical anion, making these low-frequency modes a marker for the initial electron step in the protein environment. Our results reveal how vibronic coupling influences the excited-state dynamics in the *Er*Cry4a protein system, with implications for the quantum yield of the first electron-transfer step and – thus – radical pair formation.

## 2. Results

### 2.1 Transient absorption spectra of *Er*Cry4a

European Robin cryptochrome, *Er*Cry4a protein was expressed using a previously reported method.[25] To record the steady-state absorption spectrum, the protein was dispersed in a Tris buffer solution (20% glycerol, 200 mM NaCl) at pH 8. The absorption spectrum of *Er*Cry4a (Fig. 1a, black line) shows the $S_0 \rightarrow S_1$ transition of the FAD cofactor at 2.78 eV (~ 450 nm) [62-67, 75, 76] with vibronic substructure with energy spacing of 160 meV.[77, 78] The substructure is substantially more pronounced than for free FAD in water (Figure 1a, gray line). The emission spectrum of FAD[43, 49, 79] (Fig 1a, red line) display a red-shifted, structureless emission band centered around 540 nm (2.26 eV).

We record differential transmission ($\Delta T/T$) spectra of *Er*Cry4a in buffer at a temperature of 4°C. For this we use 12-fs pump pulses centered at 2.61 eV (475 nm) and a broadband white-light continuum laser as the probe (Fig. S1). The pump pulses (shaded blue area in Fig. 1a) are tailored to overlap with the low-energy vibronic bands in the absorption spectra (black line). The time resolution of the setup is sufficient to probe vibrational coherences with frequencies of up to 3000

cm$^{-1}$.[80] Due to the low molar extinction coefficient of FAD (11300 M$^{-1}$ cm$^{-1}$)[81, 82], the $\Delta T/T$ spectra contain substantial contributions from cross-phase modulation (XPM) during and around the pump-probe overlap and other contributions of the Tris buffer.[83, 84] These signals are quantified in independent reference measurements (see Supplementary Information, Section 3) and subtracted from the data. The corrected $\Delta T/T$ is plotted in Fig. 1b as a function of the pump-probe delay, $t_d$, and the probe energy. Several features are important: (i) a long-lived ground state bleaching (GSB) band in the energy range 2.5 – 2.95 eV, (ii) a red-shifted stimulated (SE) around 2.1 – 2.35 eV, decaying within ~400 fs and (iii) three persistent excited-state absorption (ESA) signals at energies below 2.1 eV, around 2.43 eV, and above 3.0 eV. These bands have also been observed in previous studies of FAD bound to *Er*Cry4a[43] and other proteins[49, 72]. The time resolution of the present experiments allows us to discern two new spectral components: (iv) A pronounced positive band around 2.71 eV around $t_d = 0$ fs showing substructure with a splitting of ~140 meV (blue line in Fig. 1c). Its line shape resembles that of the steady-state absorption with an apparent red shift of ~65 meV. This band vanishes within 50 fs (black line in Fig. 1e). (v) Concurrently, the SE at 2.25 eV shows a finite build up, evolving from a broad ESA band at $t_d = 0$ fs to the smooth SE band after ~50 fs (Fig. 1d). A similar evolution of the early-time $\Delta T/T$ spectrum has recently been observed for FAD in water[57] and has been connected to nonadiabatic couplings between the bright $S_1$ state and the close-lying dark states.[66] $\Delta T/T$ dynamics at selected probe energies are depicted in Fig. 1e. The plot shows the sub-50-fs dynamical evolution of the spectrum and a slower decay of the SE signal at 2.25 eV, over ~ 400 fs (blue line). This slower decay monitors the electron transfer from the nearest Trp residue in the Trp chain to the FAD chromophore.[43, 55, 72]

A global analysis[85] of the $\Delta T/T$ spectra confirms this assignment (Fig. 1f). The spectra are well decomposed into three decay-associated difference spectra (DADS) with decay times of 16 fs, 360 fs and 80 ps, respectively. The slowest component (blue) reflects a quasi-static contribution $\Delta T/T$ arising from a superposition of persistent GSB of $FAD_{ox}$, matching the static absorption spectrum, and long-lived ESA of $TrpH^{\bullet+}$ and $FAD^{\bullet-}$ radicals.[43, 72, 86] These cations are formed upon electron transfer from Trp to FAD. Their plateau-like $TrpH^{\bullet+}$ absorption extends from 1.9 to 2.4 eV and is essentially independent of the position of the cation in the tetradic Trp chain.[43, 67] Hence, this $TrpH^{\bullet+}$ absorption persists during the lifetime of the [$FAD^{\bullet-}$ $TrpH^{\bullet+}$] radical pair. The finite transfer time $\tau_{ET} = 0.36$ ps between the Trp residue nearest to FAD and the chromophore is deduced from the second DADS (red curve in Fig. 1f)[43]. It shows, again, the ESA spectrum of the $TrpH^{\bullet+}$ radical cations, now with opposite sign reflecting the build-up of the signal with $\tau_{ET}$. In addition, it contains the SE from the excited state $FAD^*_{ox}$. with positive sign and centered around 2.25 eV. More details about the reconstruction of the individual species spectra from DADS are given in Refs. 43 and 87. The first DADS with 16-fs time constant (black line in Fig. 1f) contains new information about the relaxation dynamics on the excited state potential of $FAD^*_{ox}$. It shows that SE from $FAD^*_{ox}$ initially appears as a vibronically structured band around 2.7 eV. Initially, a broad negative ESA band around 2.2 eV masks the low-energy SE at 2.25 eV. This ESA decays with 16 fs, while the low-energy SE band around 2.25 eV builds up concurrently. The data thus gives evidence for a transient red shift of SE from 2.7 eV to 2.25 eV on this very fast 16-fs time scale. Similar signatures for a rapid build-up of red-shifted SE have been found for FAD in water.[57] For FAD, the red-shifted SE persists for several ps, limited by an intramolecular ET in the stacked FAD conformer.

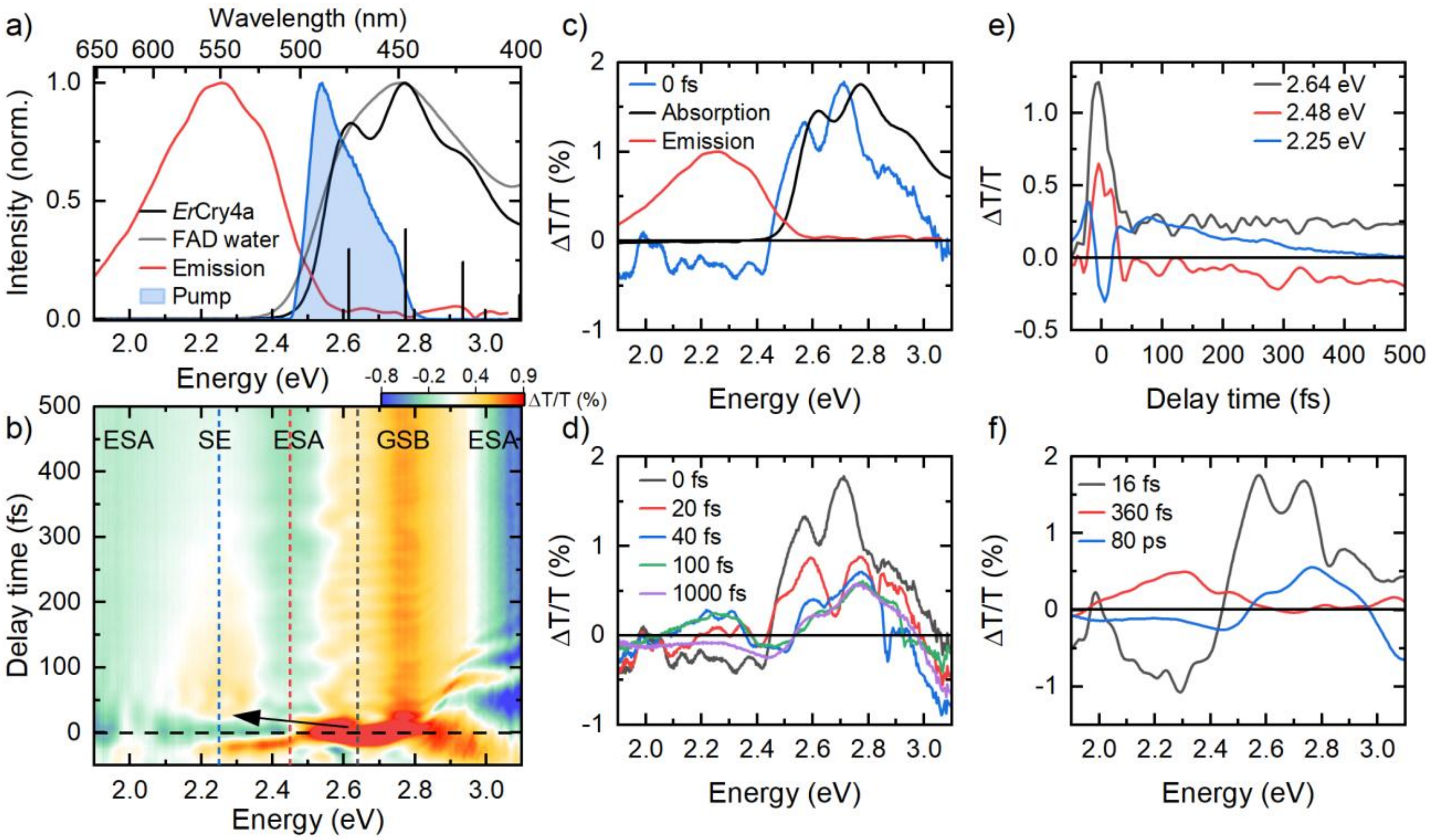

**Figure 1**. Ultrafast spectroscopy of European Robin Cryptochrome (*Er*Cry4a) protein dispersed in Tris buffer. (a) Steady state absorption of fully oxidized FAD in *Er*Cry4a (black) and in water (gray) together with the FAD emission spectrum[43] (red). The vertical black line shows a qualitative stick spectrum of FAD assuming vibronic coupling to a single vibrational mode of 160 meV energy. The pump pulse spectrum, centered at 2.6 eV, is depicted in shaded blue. (b) $\Delta T/T$ map of $Er$Cry4a as a function of probe energy and pulse delay $t_d$. Relevant ground-state bleaching (GSB), stimulated emission (SE), and excited-state absorption (ESA) transitions are marked. The black arrow indicates the rapid red-shift of the $\Delta T/T$ signal (~400 meV) during the first 50 fs. (c) Comparison of $\Delta T/T$ spectra at zero delay (blue) to the scaled steady-state absorption and emission spectra from (a). (d) $\Delta T/T$ spectra at selected delays showing the decay at ~2.6 eV and build-up of low-energy SE signal at 2.25 eV. (e) $\Delta T/T$ dynamics at selected probe energies marked by dashed vertical lines in (b), emphasizing the rapid early-time evolution and the decay of the SE at 2.25 eV on a ~400-fs time scale. This decay monitors the electron transfer from the nearest Trp to the FAD chromophore. (f) Decay-associated difference spectra with lifetimes of 16 fs (black), 360 fs (red) and 80 ps (blue), obtained from a global analysis.

## 2.2 Coupling to high-frequency vibrational modes

The high time resolution of our experiments not only allows us to track the rapid spectral evolution of the stimulated emission spectrum of *Er*Cry4a but also makes the experiments sensitive to coherent vibrational wavepacket motion in the protein environment. To probe these wavepackets, we record $\Delta T/T$ spectra and analyze the underlying incoherent pump-probe dynamics using global analysis. These incoherent transients are then subtracted from the data to obtain a map of residuals as a function of probe energy and time delay. The resulting residuals for *Er*Cry4a are depicted in Fig. 2a. Notably, persistent oscillations with periods in the 20-30 fs range appear between 2.5 and 2.9 eV, overlapping the GSB spectrum depicted as a black line in Fig. 2c. These high-frequency oscillations persist for up to 2 ps. Several lower frequency oscillations are observed between 2.2 and 2.7 eV. We first analyze the high-frequency oscillations in the Fourier transform (FT) spectrum of these residuals (see section 3, supporting information). At least five different modes with frequencies between 1150 and 1450 $cm^{-1}$ are resolved (Fig. 2b).

TA measurements are performed under comparable conditions for FAD in water and for the Tris buffer and analyzed accordingly. The FT spectra of the Tris buffer, spectrally integrated over all probe energies, are plotted in Fig. 2d as a blue line. They show sharp peaks reflecting the coherent excitation of persistent ground state vibrations by impulsive stimulated Raman scattering (ISRS).[88] The buffer modes mainly arise from glycerol, making up 20% of the buffer.[89] The integrated FT spectra of *Er*Cry4a in buffer and FAD in water, are shown as black and red lines in Fig. 2d, respectively. Buffer modes are absent, supporting the chosen XPM-correction. Several high-frequency modes are seen between 1000 $cm^{-1}$ and 1700 $cm^{-1}$, with frequencies of *Er*Cry4a that match those of FAD in water.[56, 57, 90-92] A direct comparison of the mode frequencies is presented in the Supporting Information (Table S1).

The amplitudes of the FT spectra of two pronounced high-frequency modes of *Er*Cry4a, 1228 $cm^{-1}$ and 1348 $cm^{-1}$, are plotted as a function of the probe energy as black lines in Fig. 2e and f, respectively. These vibrations correspond to mixed ring stretching and bending modes of the isoalloxazine core[56]. The striking similarity between these spectra and those for FAD in water (red line) is a clear signature that these high-frequency vibrations are only weakly affected by the protein environment.

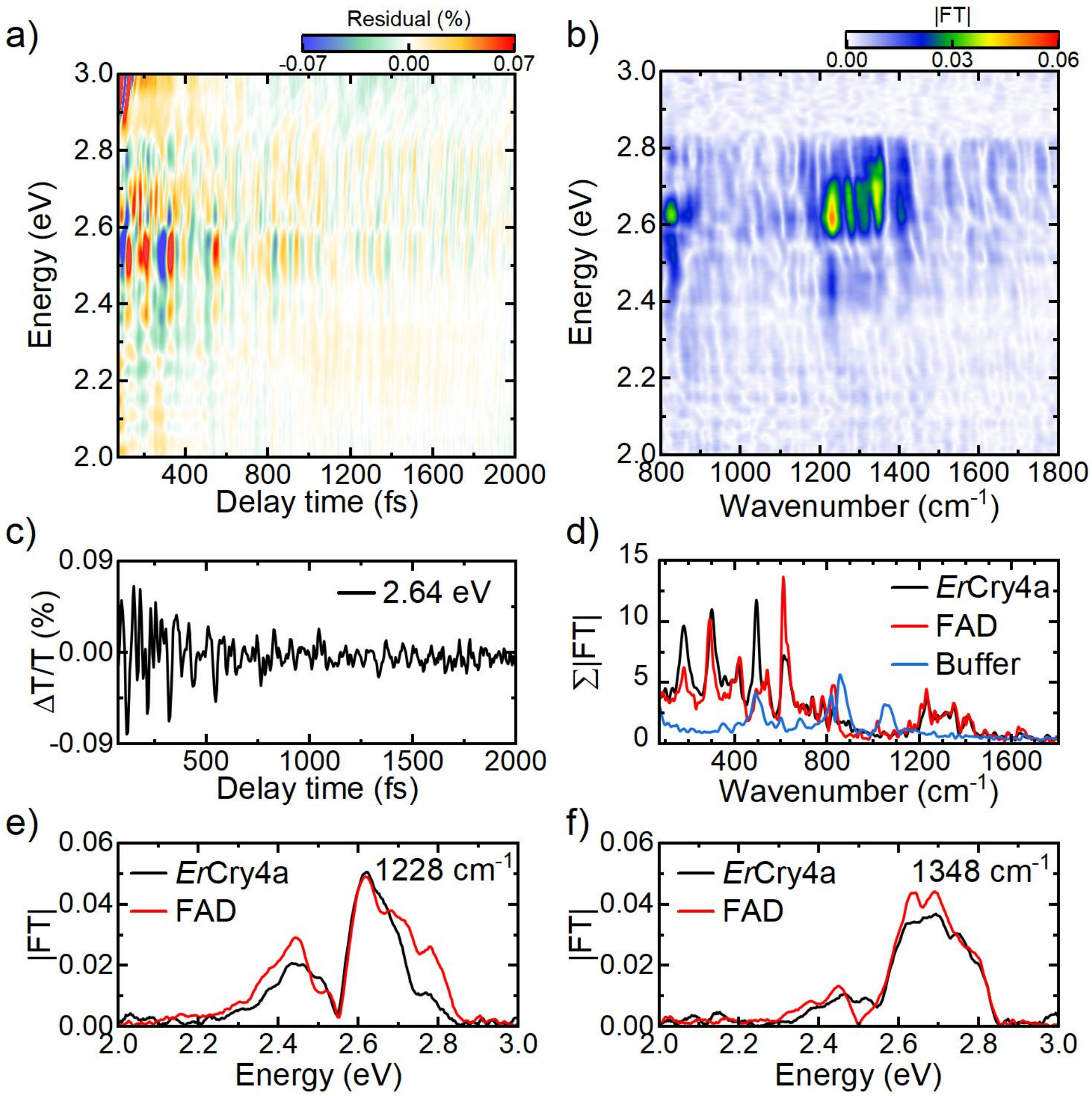

**Figure 2.** Fourier transform (FT) analysis of coherent oscillations in the $\Delta T/T$ spectra of *Er*Cry4a. a) Residual oscillations in $\Delta T/T$ as a function of the time delay $t_d$ and probe energy. b) FT spectra of high-frequency vibrational modes in *Er*Cry4a deduced from (a). c) Dynamics of the residuals in *Er*Cry4a at a probe energy of 2.64 eV. Low- and high-frequency oscillations persist for several 100 fs. d) Comparison of the amplitude of the FT spectra deduced from residual maps and integrated over the probe energy for *Er*Cry4a (black), FAD in water (red) and the Tris buffer (blue). e, f) Probe energy dependence of the mode profiles of two selected high frequency modes at 1228 cm$^{-1}$ (e) and 1348 cm$^{-1}$ (f).

The FT spectra of these two modes extend from 2.35 to 2.85 eV, covering the GSB region of *Er*Cry4a (Fig. 2b). The FT amplitude vanishes in the region of the red-shifted SE at 2.25 eV. This strongly suggests that these high-frequency oscillations are induced by wavepacket motion in the $S_0$ electronic ground state.[57] High-frequency excited state vibrations are so rapidly damped that they cannot be clearly resolved in the FT spectra. This conclusion is strongly supported by simulations based on a phenomenological 2-dimensional displaced harmonic oscillator (DHO) model that we reported for free FAD.[57, 93, 94] Also earlier quantum dynamics simulations for riboflavins[66] agree with our experimental findings. These calculations demonstrate that this rapid damping of high-frequency modes is due to nonadiabatic couplings between the optically bright $S_1$ state and the energetically close-lying dark states, mediated by high-frequency carbon backbone modes of the isoalloxazine moiety. These nonadiabatic couplings induce rapid intramolecular vibrational energy redistribution processes in the excited state of FAD that are responsible for the red shift in the SE within 50 fs. The striking similarity of the FT spectra of FAD and *Er*Cry4a in Figs. 2e and 2f indicates that these nonadiabatic couplings are similarly pronounced in water and in the protein environment.

### 2.3 Coupling to low-frequency vibrational modes

In stark contrast to the high-frequency modes, low-frequency, < 800 $cm^{-1}$, coherent oscillations in the $\Delta T/T$ data are substantially different in free FAD and in the protein environment. To quantitatively analyze these differences, we perform experiments with uncompressed probe pulses that are stretched in time to more than 200 fs. This slightly reduces the temporal resolution, yet removes higher order chirp from the pulses, facilitating the analysis of the coherent oscillations.[57, 80] The $\Delta T/T$ maps are corrected for buffer-induced XPM and probe-energy dependent shifts in time-zero. The Fourier transforms of residual maps, obtained by subtracting incoherent dynamics, are shown in Fig. 3b for *Er*Cry4a.

Several low-frequency oscillations are seen in a narrower range of probe energies around 2.6 eV. Their frequencies mainly cover the range from 150 to ~600 $cm^{-1}$ (vibrational periods of 220 to 55 fs), as shown in the FT map in Fig. 3b. Six dominant vibrational modes with frequencies of 179, 299, 429, 494, 532, and 600 $cm^{-1}$ are resolved. Interestingly, the mode profiles are spectrally sharp along the probe energy, with widths of <100 meV, much narrower than, e.g., the linear absorption or GSB spectrum. In the SE region, below 2.4 eV, the FT amplitude of all low-frequency modes is much reduced. Such FT maps have been recorded under comparable experimental conditions for FAD in water[57] and are shown in Fig. S10 of the Supporting Information. For FAD, only five peaks are clearly resolved, with frequencies of 185, 285, 429, 518, and 607 $cm^{-1}$. Some of these frequencies are shifted from those in free FAD, as also seen in Fig. 2d.

In the SE region, the FT maps of *Er*Cry4a and free FAD (Fig. S10b) are substantially different. In free FAD, several persistent low-frequency oscillations are resolved in the SE region, which are absent in *Er*Cry4a. This marked difference is emphasized by plotting, in Fig. 3c, residuals at probe energies of 2.6 and 2.25 eV for *Er*Cry4a (black) and free FAD (red).

Figures 3d and e compare the probe energy dependence of the FT amplitude for two selected modes with frequencies of 429 and 600 $cm^{-1}$. Calculations of infrared spectra (Figure S14 of the Supplementary Information) assign these vibrations to in-plane and out-of-plane deformation modes of the isoalloxazine ring, respectively. In both samples, the spectra are largely dominated by peaks around 2.6 eV. While the probe energies of the peak maxima are similar in both samples, cross sections of the *Er*Cry4a peaks along the probe energy axis are significantly narrower. Importantly, for FAD in water, distinct peaks in the FT spectra appear in the SE region around 2.3 eV for both modes. These peaks are absent in *Er*Cry4a. These peaks arise because coherent low-frequency vibrational wavepacket motion on the excited state surface periodically modulates the SE to the ground state.[57]

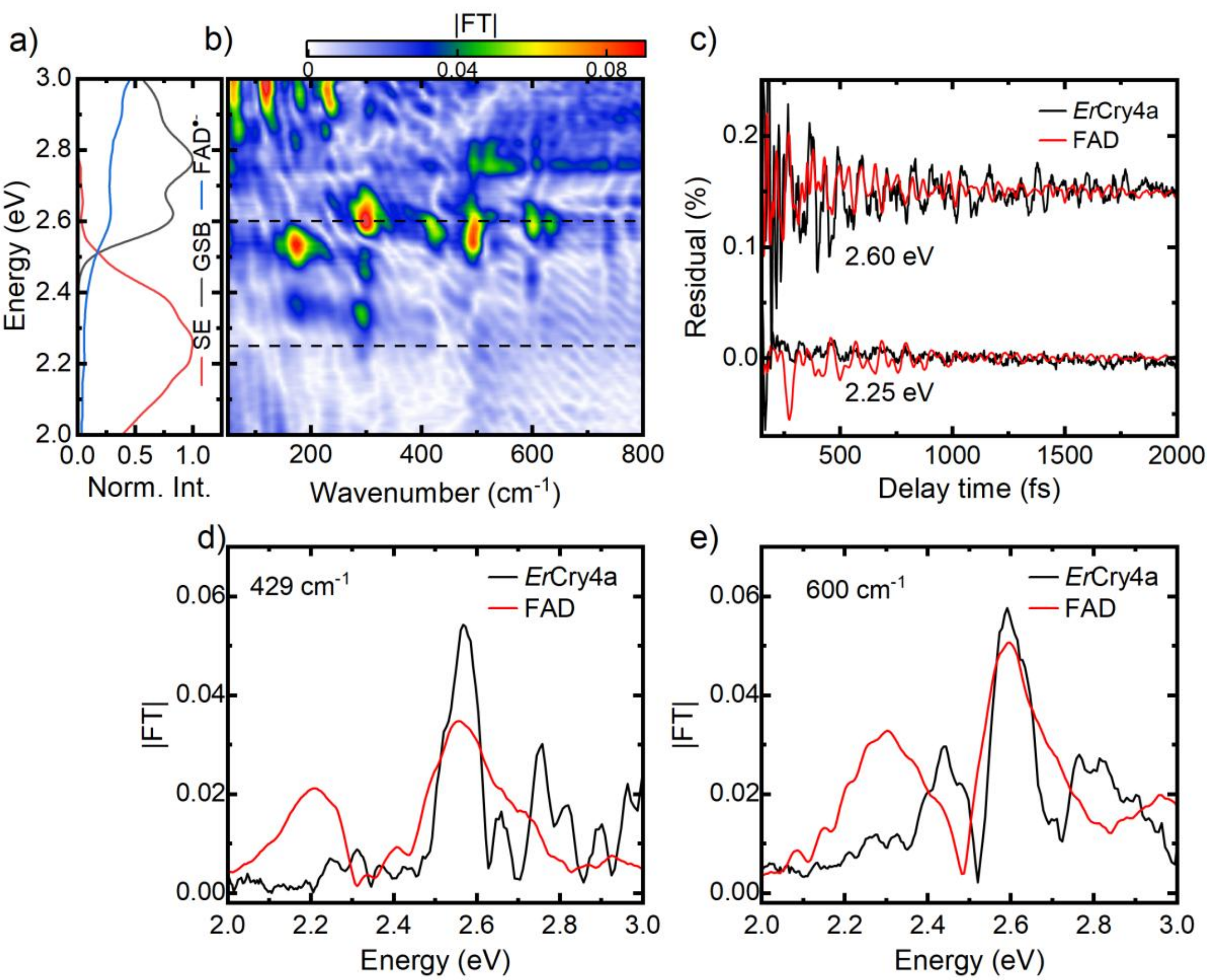

**Figure 3**. Fourier analysis of low frequency modes in *Er*Cry4a. a) Steady-state GSB (black) and SE (red) spectra of oxidized FAD in water and steady-state absorption spectrum of the semiquinone $FAD^{\bullet-}$ radical.[67] b) FT amplitude map of the low-frequency modes of *Er*Cry4a plotted as a function of probe energy. The peaks around 2.6 eV are assigned to ESA of the $FAD^{\bullet-}$ radical. Note the absence of mode amplitude in the SE region around 2.25 eV. The dotted lines show the probe energy position of the crosscut in panel c). c) Dynamics of the residuals of *Er*Cry4a (black) and FAD in water (red) at two selected probe energies of 2.6 eV and 2.25 eV. Low-frequency oscillations at 2.25 eV are present for free FAD but suppressed in the protein. d, e) Amplitude profile for two selected low-frequency modes at 429 $cm^{-1}$ (d) and 600 $cm^{-1}$ *Er*Cry4a (black) and FAD in water (red).

In the protein environment, this SE is rapidly quenched due to the ultrafast transfer of an electron from Trp to the ground state of $FAD_{ox}$. This ground state filling blocks the SE with 360 fs time constant. Therefore, coherent oscillations associated with these modes are rapidly damped and largely reduced in amplitude in the FT spectra of *Er*Cry4a.

We conclude from these observations that low-frequency modulations in the $\Delta T/T$ spectra of *Er*Cry4a cannot be related to an excited state wavepacket motion that is probed via SE transitions. They also cannot be assigned to ground state vibrations sensed via GSB because our pump pulses are so short that the amplitude of low-frequency wavepackets induced by ISRS is low. In addition, the damping time of the low-frequency oscillation in the region around 2.6 eV of more than 1 ps (Fig. 3c) is much longer than the ET time. This indicates that these oscillations reflect coherent wavepacket motion in the $FAD^{\bullet-}$ radical anion that is formed upon ET. The oscillations are sensed via ESA of the formed radical anion. A coarse comparison of the peak energies in the FT map of *Er*Cry4a with a simulated absorption spectra of $FAD^{\bullet-}$[67], depicted in Fig. 4a as a blue line, and of the riboflavin radical anion (Fig. S13 of the Supplementary Information) show that optical absorption around 2.6 eV in both FAD and $FAD^{\bullet-}$, is generally consisten with our observations. We therefore suggest that low-frequency coherent wavepacket motion on the excited $FAD^{*}_{ox}$ state is transferred to the $FAD^{\bullet-}$ radical anion during electron transfer. It is read out via ESA from either $FAD^{*}_{ox}$ or from the anion surface after ET. The low-frequency coherence persists sufficiently long (~ 1 ps) to give rise to the narrow FT peaks that are seen in Fig. 3b.

It is tempting to assign the additional FT peak at 494 $cm^{-1}$, observed only in case of *Er*Cry4a as shown in Fig. 3b and 2d, to a vibrational mode in $FAD^{\bullet-}$. Such a mode, however, is not seen in the Raman spectra of $FAD^{\bullet-}$.[91] We have therefore searched for an independent marker for persistent vibrational coherence in the $FAD^{\bullet-}$ anion. Since the simulations (Fig. S13) suggest a pronounced absorption around 3.2 eV, we have performed TA experiments with a white light continuum as a probe that extends to 3.5 eV.[43] For these experiments, spectrally narrower pump pulses around 450 nm provide a time resolution of ~30 fs, sufficient to probe vibrational motion with frequencies below 600 $cm^{-1}$. The FT maps that result from these measurements are shown for *Er*Cry4a and FAD in water in Fig. 4.

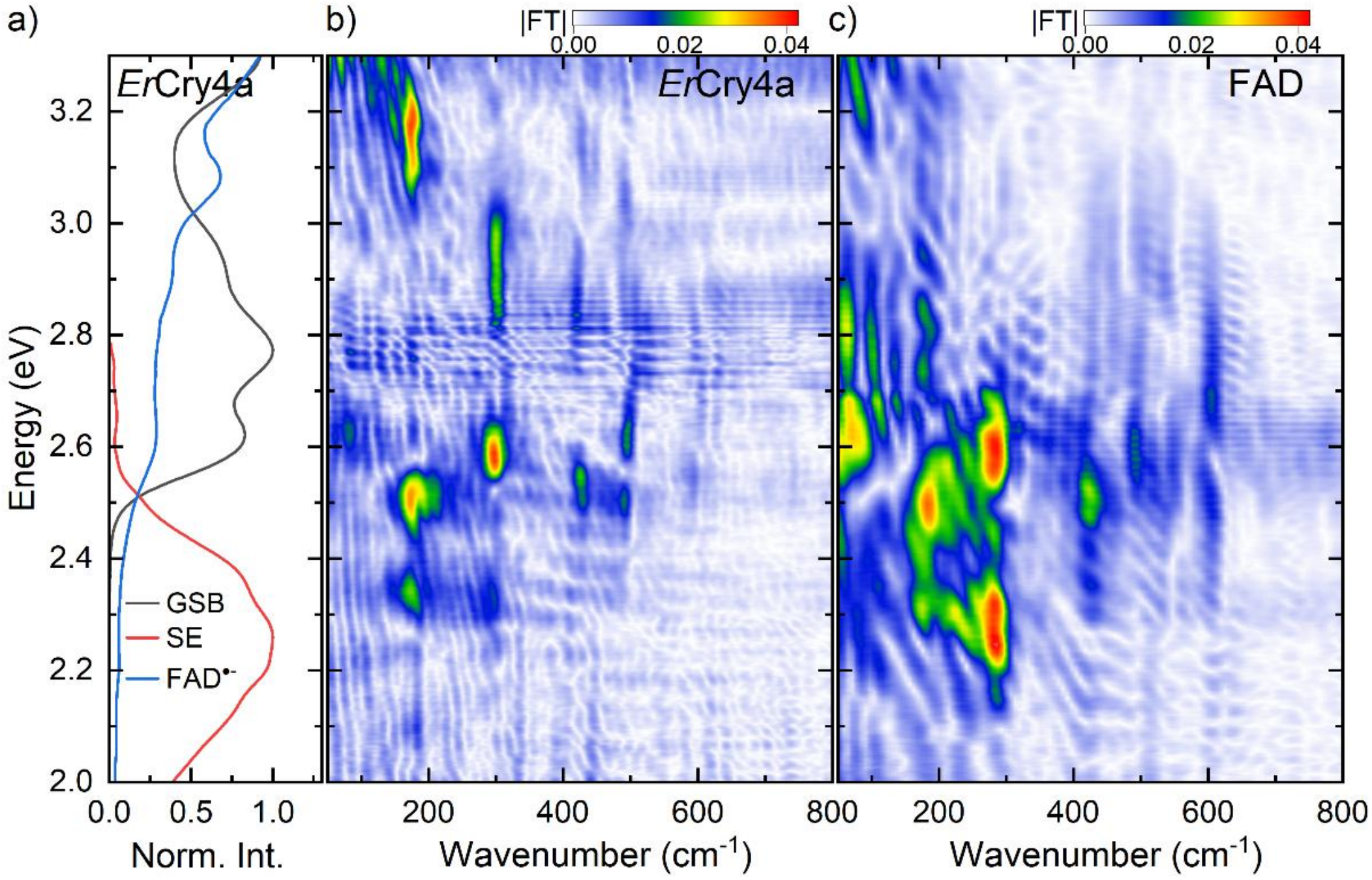

**Figure 4**. Fourier analysis of low frequency modes in *Er*Cry4a and FAD water. To identify vibrational marker bands for $FAD^{\bullet-}$ coherence, $\Delta T/T$ measurements have been performed using narrowband excitation at 450 nm, probing with a white-light continuum extending to 3.5 eV. a) Steady-state GSB (black) and SE (red) spectra of oxidized FAD in protein environment and steady-state absorption spectrum of the $FAD^{\bullet-}$ radical.[67] b) FT amplitude map of the low-frequency modes of *Er*Cry4a. c) FT amplitude map of the low-frequency modes for FAD in water. Notice the peaks at 3.15 eV for the 179 $cm^{-1}$ mode and at 2.9 eV for the 299 $cm^{-1}$ mode of *Er*Cry4a. These peaks mark low-frequency coherent vibrations in $FAD^{\bullet-}$ after photoinduced electron transfer from a nearby Trp residue to the FAD chromophore in the protein.

Importantly, peak structures obtained in the FT maps for probe energies below 2.7 eV are, for both samples, very similar to those observed and depicted in Fig. 3b for *Er*Cry4a. No additional vibrational frequencies are recorded for probe energies above 2.7 eV for FAD in water (Fig. 4c). In the case of *Er*Cry4a (Fig. 4b), instead, two pronounced new peaks appear in the FT map along the probe energy axis. The 179 $cm^{-1}$ mode peaks at ~3.15 eV and the 299 $cm^{-1}$ mode at around 2.9 eV probe energy. At both energies, we find peak structure in the $FAD^{•-}$ absorption spectrum (Fig. 4a) and in the simulated spectrum in Fig. 5a.[67] This implies that wavepacket motion in the anion can be preferentially read out by the probe laser at these energies. We take this as convincing evidence that the sharp low-frequency peaks in the FT map of *Er*Cry4a are markers for coherent vibrational wavepacket motion on the $FAD^{•-}$ radical anion potential energy surface after ET from the nearest Trp in the tetradic chain of *Er*Cry4a to the FAD chromophore.

## 2.4 Quantum chemical calculations

We then use quantum chemical calculation based on time-dependent (TD-)DFT[95, 96] as well as multireference DFT (DFT/MRCI) with the R2022 Hamiltonian, to obtain the optical spectra of neutral riboflavin (RF) and riboflavin radical anion ($RF^{•-}$), see section 3.6 Supplementary Information for more details.[97, 98] The results obtained from these simulations are depicted in Fig 5a. The deduced electronic transition energies and oscillator strengths for the neutral species (black squares) match well with the experimentally observed energies for the *Er*Cry4a. The calculations for the negatively charged species predict optical transitions slightly below the first electronic transition in the neutral state. Higher energy peaks appear in the optical absorption spectrum of $RF^{•-}$ at around 3.0 and 3.2 eV. Their energies match well with the peaks seen in the FT spectra of *Er*Cry4a in Fig. 4b for the 179 $cm^{-1}$ and 299 $cm^{-1}$ modes. Importantly, these peaks are absent in the FT spectra of FAD in water (Fig. 4c), another strong argument for their assignment to vibrational coherence on the anion surface. Optical spectra deduced using TD-DFT gives similar results, see Fig. S13 of the Supplementary Information. We further computed IR spectra of RF and $RF^{\bullet-}$, see Fig. S14 of Supplementary Information. These spectra are generally in reasonable agreement with the results of the FT spectra. These IR spectra provide a basis for assigning the low-frequency modes probed experimentally. As in the experiment, distinct changes in the IR spectrum are observed when going from the charge neutral to the anionic spectrum. Our current understanding of *Er*Cry4a photophysics that results from the presented experimental and theoretical results in summarized in Fig. 5b.

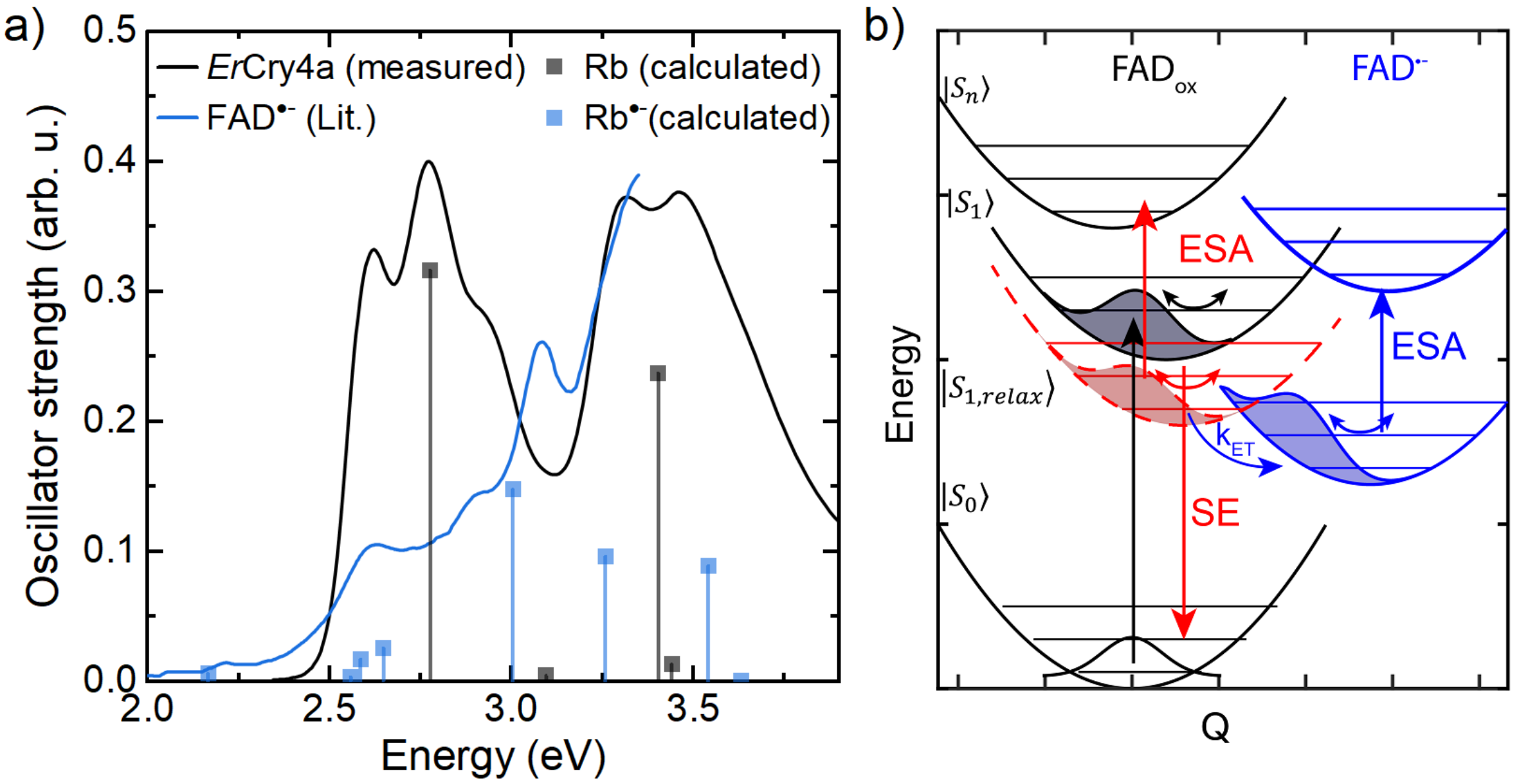

**Figure 5**. Theoretical calculations and schematic potential energy surface. a) Quantum-chemical calculations of the optical spectra of neutral riboflavin (black sticks) and riboflavin radical anion (blue sticks) compared with the experimental absorption spectrum of FAD in protein environment (*Er*Cry4a, black curve, same data as in Fig. 1a) and absorption spectrum of the $FAD^{\bullet-}$ (blue curve) from literature.[67] b) Schematic potential energy surface. Optical excitation launches a wavepacket in the $|S_1\rangle$ state which can be monitored by either SE or ESA transitions. Nonadiabatic couplings

to high-frequency modes induce rapid relaxation on the $|S_1\rangle$ surface towards a lower-energy configuration $|S_{1,relax}\rangle$. Coherent vibrational motion of different low-frequency wavepackets is preserved during this relaxation. The transfer of an electron from the nearest Trp to the FAD chromophore forms $FAD^{\bullet-}$ radical anions (blue) at a rate of $k_{ET} \simeq 2.5$ ps$^{-1}$ while preserving vibrational coherence along the low-frequency coordinates. This results in the observation of coherent wavepacket motion in the $FAD^{\bullet-}$ radical anions by ESA transitions (blue).

### 2.5 Low-frequency coherences in Mutants

To provide independent support for our conclusion that our experiments on *Er*Cry4a probe vibrational coherence in the $FAD^{\bullet-}$ state that persist after ET, we extend our experimental studies to two mutants of *Er*Cry4a, in which one of the Trp residues is replaced by a redox-inactive phenylalanine (Phe). This gives the opportunity to site-selectively block the electron transfer cascade in the protein at the position of the mutant.[43] We investigate two mutants: the mutant $W_AF$, in which the Trp residue W395 closest to FAD is replaced by Phe, and $W_BF$, in which the second Trp, W372, is replaced. Consequently, the first ET pathway is blocked in $W_AF$, yet remains open in $W_BF$.[43] $W_BF$ blocks the second electron transfer step in the Trp chain.

The FT maps deduced from these measurements are reported in Fig. 6a for the $W_BF$ mutant and in Fig. 6b for $W_AF$, respectively. The overall shape of the $W_BF$ map is in convincing agreement with the map reported for *Er*Cry4a in Fig. 4. The $W_AF$ map matches that for FAD. Indeed, the distinct $FAD^{\bullet-}$ low-frequency marker peaks at 3.15 eV for the 173 cm$^{-1}$ mode and at 2.9 eV for the 300 cm$^{-1}$ mode are present in $W_BF$, with similar amplitude to that in the *Er*Cry4a measurements, yet suppressed in $W_AF$. Thus, blocking the ET in $W_AF$ completely suppressed the coherence transfer to $FAD^{\bullet-}$. To emphasize the distinct difference in the spectra of $W_AF$ and $W_BF$, we compare, in Fig. 6c,d the probe energy dependence of the identified modes in the mutants with that in *Er*Cry4a and FAD. Selected marker bands of excited-state vibrational coherences in FAD are indicated by red Gaussian curves superimposed on the experimental data. Markers of $FAD^{\bullet-}$ coherences are depicted as black Gaussians. Importantly, the width of the $FAD^{\bullet-}$ band at 2.6 eV is significantly narrower than that of the corresponding FAD peak in *Er*Cry4a and $W_BF$.

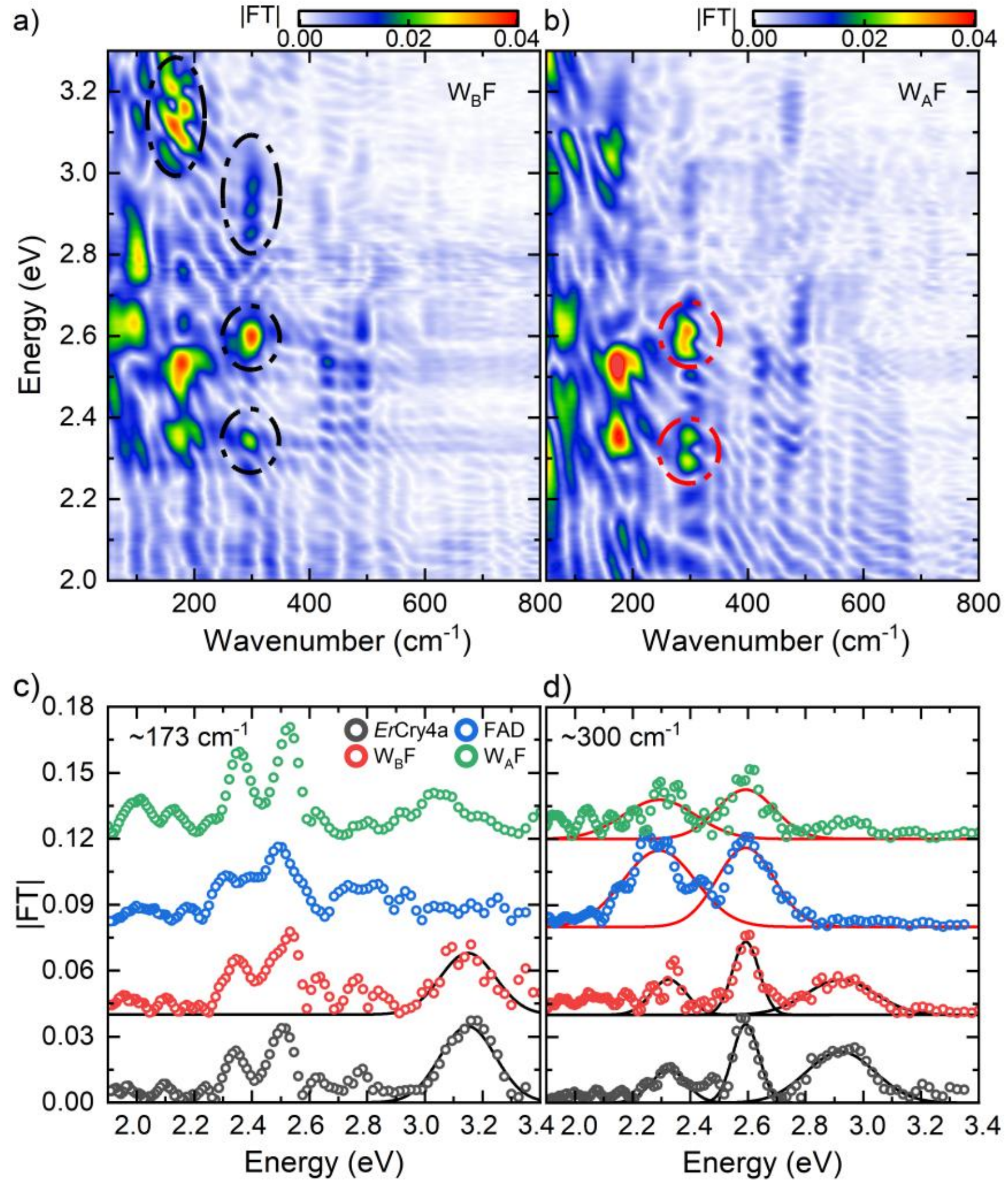

**Figure 6.** Fourier analysis of low frequency modes in *Er*Cry4a mutants. a) FT map for the mutant $W_BF$, for which ET between Trp and FAD is allowed. b) FT map for $W_AF$, blocking ET between Trp and FAD. Specific marker modes for $FAD^{\bullet-}$ are encircled in black in panel a) and marker modes for FAD are highlighted by red circles in panel b). c, d) Experimentally obtained probe energy dependencies of selected modes close to 173 cm$^{-1}$ (c), and 300 cm$^{-1}$ (d). The data from different samples are shown as black circles (*Er*Cry4a), red circles ($W_BF$), blue circles (FAD), and green circles ($W_AF$), respectively. The solid red lines are Gaussian fits acting as guides to the eye to marker bands for FAD (red) and show the absence $FAD^{\bullet-}$ when blocking the electron transfer.

## 3. Discussion

Transient absorption spectroscopy with 12-fs, sub-vibrational cycle time resolution provides insight into the primary photophysical processes in *Er*Cry4a proteins. Following impulsive blue-light excitation of the FAD chromophore, a vibronically structured SE signal emerges from the excited-state Franck-Condon region (schematically represented as $|S_1\rangle$ in Fig. 5b). This SE is only marginally Stokes-shifted by 60 meV with respect to the linear absorption. The red shift is most likely attributable to electronic solvation of the chromophore, too fast to be resolved with the present time resolution.[99] With a relaxation time of 16 fs, the band transforms into a structureless SE band, red-shifted by almost 0.4 eV with respect to the electronic 0-0 transition. This early spectral evolution matches recent observations for free FAD in water quantitatively[57] and confirms theoretical predictions for riboflavin.[66] These calculations suggest rapid IVR processes on the excited state surface of $FAD_{ox}^{*}$, induced by vibronic couplings between the bright $\pi\pi^*$and dark $n\pi^*$ states and mediated by couplings to high-frequency modes of the isoalloxazine ring.[66] On these mixed nonadiabatic surfaces, coherent high-frequency vibrational wavepackets are rapidly damped out before the system relaxes into a new quasi-equilibrium configuration ($|S_{1,relax}\rangle$). The similarity between the early time spectroscopic signatures of free FAD and *Er*Cry4a suggests that the underlying vibronic couplings and IVR processes are only weakly affected by the protein environment. In *Er*Cry4a the resulting SE is then quenched with a decay time of $\tau_{ET} = 0.36$ ps (Fig. 1f). This decay monitors the first electron transfer step in the tetradic Trp chain in the protein, the transfer of an electron from the W395 Trp residue to the FAD chromophore.[12, 43, 72] In contrast, the SE persists for several ps in free FAD in water, before it partially decays via intramolecular electron transfer in the stacked FAD conformer.[43, 48, 49] This difference is important for analyzing the vibrational wavepacket dynamics in *Er*Cry4a.

Our experiments resolve several high-frequency vibrational modes between 1000 and 1700 $cm^{-1}$, launched by ISRS in the electronic ground state of the FAD. Their oscillations persist for more than a ps and are very similar for protein-bound FAD and FAD in aqueous solution (Fig 2c and 3c). Clear signatures for such high-frequency vibrations in the excited state, however, are neither resolved for *Er*Cry4a nor for free FAD. The strong nonadiabatic couplings in the excited states dampen these high-frequency vibrations within a few oscillation cycles.[66]

Substantial differences between protein-bound and free FAD are observed in several low-frequency, excited state vibrations which coherently modulate the TA signal. For free FAD, these modes can be read out through their oscillatory modulation of the red-shifted SE band. For *Er*Cry4a, however, these signatures disappear since the SE is rapidly quenched by electron transfer from a nearby Trp to FAD. Instead, low-frequency coherent wavepacket motion in the protein environment is monitored through distinct and spectrally unexpectedly sharp ESA transitions (blue arrow in Fig. 5(b)). In the probe energy region around 2.6 eV, these ESA peaks are similar in energy to the broader peaks seen in free FAD (Fig. 3 and 4). At higher probe energies, we find distinct *Er*Cry4a ESA marker bands that are absent in free FAD. Their energies agree reasonably well with peaks in the absorption spectrum of the $FAD^{\bullet-}$ radical anion. Quantum-chemical calculations support this assignment (Fig. 5a). These observations suggest that low-frequency vibrational wavepacket motion, initiated by impulsive optical excitation of the excited state surface of $FAD_{ox}^{*}$ is coherently transferred to the $FAD^{\bullet-}$ radical anion and persists beyond the electron transfer time as depicted schematically in Figure 5b. These conclusions are strongly supported by comparative studies of two mutants of *Er*Cry4a in which this electron transfer can be site-selectively suppressed.

## Summary and conclusion

In summary, our study sheds new light on the primary photophysical processes in European robin cryptochrome 4a, the primary candidate for being the light-dependent magnetic field sensor of night-migratory birds. Our results reveal distinct signatures for pronounced nonadiabatic couplings among the lowest-lying electronically excited states of *Er*Cry4a. Vibronic coupling to high-frequency modes causes a red shift of the stimulated emission by more than 400 meV within 16 fs, quenching coherent high-frequency vibrational motion in the excited state of the protein. These nonadiabatic effects are characteristic of the FAD chromophore and are apparently only weakly affected by the protein environment.

Our experiments allow us to disentangle coherent vibrational wavepacket motion on the electronic ground- and excited state surfaces of *Er*Cry4a, and specifically to track the temporal evolution of several excited state low-frequency vibrational modes. By observing distinct markers for the excited state absorption from the $FAD^{\bullet-}$ radical anions, we provide strong evidence that their vibrational coherence is preserved beyond the electron transfer from a nearby Trp residue to the FAD chromophore. Interestingly, in the protein environment, the ESA appears in the form of several distinct and spectrally sharp peaks in the differential transmission spectrum. This shows that time-resolved vibrational spectroscopy not only monitors the transient evolution of the redox state of the charge transfer system but also partially removes the congestion and inhomogeneous broadening in the optical spectra of such proteins by selectively probing specific Franck-Condon transitions in the potential energy surface of the protein. These conclusions are confirmed by time-resolved vibrational spectroscopy of two *Er*Cry4a mutants in which electron transfer and, thus, transfer of vibrational coherence to the radical anion, is site-selectively controlled.

Our study presents clear experimental evidence that specific low-frequency vibrational coherences are preserved during the initial, ultrafast electron transfer from a nearby tryptophan to the flavin chromophore in a cryptochrome protein. This naturally raises the question of what role such nonequilibrium vibrational excitations play in the subsequent sequence of electron-transfer events along the tetradic tryptophan chain and, ultimately, in the formation of long-lived, magnetically sensitive radical pairs in cryptochromes. Our results suggest that the combination of advanced ultrafast (multidimensional) coherent spectroscopies and time-domain quantum simulations may provide qualitatively new insight into the nonadiabatic vibronic couplings[87, 100] that drive the electron transfer process in cryptochromes and, more generally, in complex molecular environments.[101, 102] In this context, the synergistic application of time-resolved spectroscopy with sub-vibrational-cycle resolution and redox-state-selective spectroelectrochemical methods[103] may prove particularly powerful.

ASSOCIATED CONTENT

**Supporting Information**.

Detailed descriptions of sample preparation, the experimental setup, data evaluation steps (including time-zero correction, XPM removal, global fitting, and Fourier analysis), as well as theoretical calculations are available in the Supporting Information. IR-mode movies illustrating the extracted vibrational motions accompany the dataset.

The following files are available free of charge.

Supplementary Information (PDF)

Vibrational modes (MPG)

AUTHOR INFORMATION

**Corresponding Author**

*Correspondence to: christoph.lienau@uni-oldenburg.de or henrik.mouritsen@uni-oldenburg.de

**Present Addresses**

†DT, present address: Fachbereich Physik, Philipps-Universität Marburg, 35032 Marburg, Germany

**Author Contributions**

K.K.: data collection and analysis, discussion and interpretation, manuscript preparation. D.T.: data collection and analysis, discussion and interpretation, review. M.K., G.D., and S.B.: Protein synthesis, review. J.P.G. and P.S.: theoretical support, manuscript preparation and review. B.G.: review. A.D.S., H.M., and C.L. conceived project, proposal writing and review. C.L.: main supervision, discussion and interpretation, manuscript development. H.M.: protein synthesis supervision. The manuscript was written through contributions of all authors. All authors have given approval to the final version of the manuscript.

ACKNOWLEDGMENT

We acknowledge funding from Deutsche Forschungsgemeinschaft, DFG, (SFB 1372 "Magnetoreception and navigation in vertebrates" (project no. 395940726), Li 580/16-1, DE 3578/3-1, JPG GO 2430/3-3 (project no. 393271229)), Germany's Excellence Strategy – EXC 3051/1 "NaviSense" – project number 533653176, and the Niedersächsische Ministerium für Wissenschaft und Kultur (DyNano and Wissenschaftsraum ElLiKo).